\documentclass[aps, prb, twocolumn, amsmath, amssymb, nofootinbib, superscriptaddress]{revtex4-2}

\usepackage{color}
\usepackage[colorlinks=true,linkcolor=blue,citecolor=blue,anchorcolor=green,urlcolor=blue]{hyperref}

\usepackage{bm}
\usepackage{xcolor}
\usepackage{amsmath}
\usepackage{amsfonts}
\usepackage{dsfont}
\usepackage{graphics}
\usepackage[caption=false]{subfig}
\usepackage{overpic}
\usepackage[capitalise]{cleveref}
\usepackage{tikz}
\usepackage{simpler-wick}
\usepackage{physics}
\usepackage{braket}
\usepackage{mathrsfs}

\newcommand{\diff}{\mathrm{d}}

\newcommand{\ee}{\mathrm{e}}

\def\be{\begin{equation}}
\def\ee{\end{equation}}

\begin{document}
\title{Multipolar Ferroelectricity in the Mott Regime}
\author{Pengwei Zhao}
\thanks{These authors contributed equally to this work.}
\affiliation{International Center for Quantum Materials, School of Physics, Peking University, Beijing 100871, China}
\author{Jiahao Yang}
\thanks{These authors contributed equally to this work.}
\affiliation{International Center for Quantum Materials, School of Physics, Peking University, Beijing 100871, China}
\author{Gang v.~Chen}
\email{chenxray@pku.edu.cn}
\affiliation{International Center for Quantum Materials, School of Physics, Peking University, Beijing 100871, China}
\affiliation{Collaborative Innovation Center of Quantum Matter, 100871, Beijing, China}
\date{\today}

\begin{abstract}
 Ferroelectricity has been one major focus in modern fundamental research and technological application. 
 We consider the physical origin of improper ferroelectricity in Mott insulating materials. 
 Beyond the well-known Katsura-Nagaosa-Balatsky’s inverse Dzyaloshinskii-Moriya mechanism 
 for the noncollinearly ordered magnets, we point out the induction of the electric polarizations 
 in the multipolar ordered Mott insulators. Using the multiflavor representation 
 for the multipolar magnetic moments, 
 we can show the crossover or transition from the pure inverse Dzyaloshinskii-Moriya mechanism    
 to the pure multipolar origin for the ferroelectricity, 
 and also incorporate the intermediate regime with the mixture of both origins. 
 We expect our results to inspire the reexamination of the ferroelectricity 
 among the multipolar-ordered magnets.
\end{abstract}

\maketitle

\section{Introduction}
The origin of electric polarization in solid-state systems 
has been an enduring subject in modern condensed matter physics. 
Even for the non-interacting band insulators, this question turns out to be quite fundamental. 
It is realized that the electric polarization of the band insulators 
in the quantum case is actually multi-valued, 
and thus, is related to some kind of phase variables~\cite{baroniGreensfunctionApproachLinear1987, king-smithTheoryPolarizationCrystalline1993, restaMacroscopicElectricPolarization1993, restaTheoryPolarizationModern2007, spaldinBeginnersGuideModern2012}. The progress was made until the introduction of the Berry phase effects for the Bloch states of the band electrons, and it was pointed out that the electric polarization is related to the integration of the Berry phase for the Bloch electrons over the Brillouin zone \cite{bernevigTopologicalInsulatorsTopological2013, vanderbiltBerryPhasesElectronic2018}. This result and understanding are revived in the era of (magnetic) topological insulators where the quantized axion magnetoelectric response was discovered~\cite{nennoAxionPhysicsCondensedmatter2020}. 
In the Mott insulating regime, the degrees of freedom are localized spins and orbitals 
instead of the physical electrons. Based on the irrelevance of the electron bands, 
it seems that the relation to the electron Berry phase 
does not seem to be directly applicable to the Mott regime. 
The emergence of improper ferroelectricity with 
the magnetic degrees of freedom in the Mott insulators 
has been an interesting subject for multiferroics, 
where the electric polarization is considered as 
the outcome of magnetism~\cite{bulaevskiiElectronicOrbitalCurrents2008, eerensteinMultiferroicMagnetoelectricMaterials2006, jiaBondElectronicPolarization2006, katsuraSpinCurrentMagnetoelectric2005, mostovoyFerroelectricitySpiralMagnets2006, murakawaFerroelectricityInducedSpinDependent2010, sergienkoRoleDzyaloshinskiiMoriyaInteraction2006, tokuraMultiferroicsSpinOrigin2014, windschFreemanSchmidMagnetoelectric1976,cheongMultiferroicsMagneticTwist2007, delaneySuperexchangeDrivenMagnetoelectricityMagnetic2009, mostovoyMultiferroicsDifferentRoutes2024, mostovoyTemperatureDependentMagnetoelectricEffect2010, mostovoyTheoryElectricPolarization2011, neatonFirstprinciplesStudySpontaneous2005, spaldinAdvancesMagnetoelectricMultiferroics2019, spaldinRenaissanceMagnetoelectricMultiferroics2005, vanakenOriginFerroelectricityMagnetoelectric2004}. 
It is expected that the magnetoelectric coupling in the multiferroic materials 
could enable the electric control of magnetism, and vice versa \cite{Massarelli2019OrbitalEdelstein,liWritingDeletingSkyrmions2021}. 
One well-known mechanism for the improper ferroelectricity in the Mott insulators 
is the Katsura-Nagaosa-Balatsky’s inverse Dzyaloshinskii-Moriya mechanism that 
was proposed by connecting to the spin current for the non-collinearly ordered magnets~\cite{katsuraSpinCurrentMagnetoelectric2005, sergienkoRoleDzyaloshinskiiMoriyaInteraction2006}. 
Despite its success, the origin of the ferroelectricity in the Mott regime 
is still not a fully resolved problem. Within the limit of our understanding, 
we think there are at least two more important aspects of this problem, and we explain below.

The first aspect is about the local moment structure
in the Mott regime. Often, the local moment is not simply the magnetic dipole moment 
with a pure spin-$S$ contribution and could involve high-rank magnetic moments, 
such as magnetic quadrupole and octupole moments \cite{banerjeeMultipolarMultiferroics4d22024}. 
In the original inverse Dzyaloshinskii-Moriya mechanism, only the magnetic dipole moment is considered \cite{katsuraSpinCurrentMagnetoelectric2005}. 
It is thus natural to explore the relation between the electric polarization and the high-rank 
magnetic multipoles. 
The second aspect is about the charge fluctuations especially 
since the ferroelectricity is related to the charge degrees of freedom. 
While the charge fluctuations are suppressed in the strong Mott regime, 
they are quite significant in the weak Mott regime \cite{Lee2005U1,Motrunich2006Orbital,Motrunich2005Variational}. 
Although the mechanisms for the ferroelectricity extend to the weak Mott regime, 
the strong charge fluctuations in the weak Mott regime could induce new mechanisms 
for ferroelectricity. Moreover, the physical spin in the weak Mott regime  
is not very far from the electron in the metallic side, and thus, 
one may expect the Berry phase physics to extend to the weak Mott regime 
when the system is in certain spin liquid phases. In this work, 
we focus our study on the first aspect and come to the second aspect in a later work.

We work in the strong Mott regime with multipolar order. 
In particular, for the specific $J = 1$ case, it is narrowed down as the quadrupolar ferroelectricity, and other multipolar ferroelectricity 
such as the octupolar one has been considered in Ref.~\cite{banerjeeMultipolarMultiferroics4d22024}.
We then explore the relationship between the electric polarization and the local magnetic moment 
and single out the contribution from the quadrupole moment. This mechanism is beyond the well-known 
inverse Dzyaloshinskii-Moriya mechanism, and is quoted as ``multipolar ferroelectricity''. 
In the actual formulation of our calculation for ${J = 1}$, we are able to obtain the contribution 
from both the dipole and quadrupole channels, and can continuously vary from 
the pure inverse Dzyaloshinskii-Moriya mechanism to the pure quadrupolar ferroelectricity. 

The remainder of the paper is organized as follows.
Section~\ref{sec:Model} introduces the spin-orbit coupled states for the effective moment $J=1$. An analysis of magnetic dipoles and quadruples carried by these states is also given in this section.
In section~\ref{sec:Three-site cluster}, we consider a minimal three-site cluster model to illustrate the fundamental mechanism of multipolar ferroelectricity.
After providing the analytical expression for the electric polarization in Sec.~\ref{sec:Electric polarization}, 
we summarize the paper with discussion and conclusion in Sec.~\ref{sec:Discussion}.

\section{Single Ion Physics}
\label{sec:Model}
\subsection{Ground States}
\begin{figure}[t]
    \centering
    \includegraphics[width=0.95\linewidth]{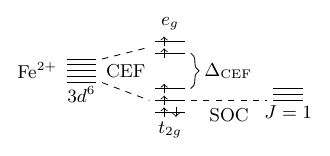}
    \caption{ 
    A schematic diagram of the level splitting of the $3d$ orbitals of the Fe$^{2+}$ ion
    in the octahedral environment. 
    The crystal electric field (CEF) splits the fivefold $3d$ orbitals 
    into the $e_g$ and $t_{2g}$ manifolds with an energy gap $\Delta_{\text{CEF}}$. 
    Arrows show the electron spin configuration on $e_g$ and $t_{2g}$ orbitals.
    Within the $t_{2g}$ subspace, the spin-orbit coupling of ${L=1}$ and ${S=2}$ 
    produces a new ground manifold with the total angular momentum ${J=1}$.}
    \label{fig:level_splitting}
\end{figure}

The spin-orbit-entangled ${J=1}$ local moment can arise from a magnetic ion, 
such as Fe$^{2+}$, which has a 3$d^6$ electron configuration in an octahedral crystal field.
In this environment, the crystal field splits the fivefold degenerate $d$-orbitals into 
a $t_{2g}$ triplet as the ground state and an $e_g$ doublet, 
separated by an energy gap 
$\Delta_{\text{CEF}}$~\cite{fazekasLectureNotesElectron1999a,maekawa2004PhysicsTransitionMetal}.
Following the Hund's first rule, we consider the high-spin configuration for 3$d^6$, 
with ${S=2}$ and an electron distribution of $t_{2g}^4e_g^2$. 
As the $t_{2g}$ shell is partially filled, there exists a three-fold 
degeneracy for the orbital configuration, and the spin-orbit coupling (SOC) 
is active in the linear order. The total orbital angular momentum 
in this degenerate manifold is equivalent to an orbital moment $\bm{L}$ with ${L=1}$.
After including the SOC $H_{\text{SOC}}=\lambda\bm{L}\cdot\bm{S}$ ($\lambda>0$), 
the total angular momentum ${\bm{J}=\bm{L}+\bm{S}}$ is used to label the local moment, 
and it has a threefold degenerate ground state manifold with ${J=1}$ for the Fe$^{2+}$ ion \cite{chenQuadrupoleMomentsTheir2023,baiHybridizedQuadrupolarExcitations2021}
(see \cref{fig:level_splitting}).

In the following, the wavefunctions of the ${J=1}$ moment
are constructed explicitly, which is necessary for the later 
investigation of ``multipolar ferroelectricity''. To begin with, 
we derive the effective orbital angular momentum $\bm{L}$ from 
the hole representation since six electrons are more than half-filling.
Thus, the orbital configuration of four holes is $t_{2g}^2e_g^2$-like with three possibilities,
\begin{equation}
\begin{split}
    & \ket{a}=A^\dagger|0\rangle=d^\dagger_{3z^2}d^\dagger_{x^2-y^2}d_{zx}^\dagger d_{xy}^\dagger\ket{0}, 
    \\
    & \ket{b}=B^\dagger|0\rangle=d^\dagger_{3z^2}d^\dagger_{x^2-y^2}d_{xy}^\dagger d_{yz}^\dagger\ket{0}, 
    \\
    & \ket{c}=C^\dagger|0\rangle=d^\dagger_{3z^2}d^\dagger_{x^2-y^2}d_{yz}^\dagger d_{zx}^\dagger\ket{0}, 
\end{split}
\label{eq:os_ABC}
\end{equation}
where $\ket{0}$ is the vacuum state of holes (full state of electrons)
and $d^\dagger_{a}$ creates a hole on the $a$ orbital with 
$a={{3z^2-r^2},x^2-y^2,xy,yz,zx}$, and $3z^2$ refers to the $3z^2-r^2$ orbital. 
Here we choose the quantization of $\bm{L}$ along $z$-direction.
Based on the states in Eq.~\eqref{eq:os_ABC}, we can construct 
the corresponding orbital angular momentum operators $\bm{L}$ and the eigenstates of $L_z$ as
\begin{equation}
\begin{split}
& \ket{L_z=\pm 1}=\frac{1}{\sqrt{2}}[\ket{a} \pm i\ket{b}], \quad \ket{L_z=0}=\ket{c}. 
\end{split}
\end{equation}
Regarding the spin-orbit coupling, the eigenstate of total angular momentum 
$\ket{J,J_z}$ is expressed in the decoupled representation of $\ket{L,L_z,S,S_z}$ 
through Clebsh-Gordan coefficients. This leads to three-fold eigenstates with total angular momentum $J=1$. The explicit forms in terms of individual holes 
are listed in the App.~\ref{app:spin configuration}. 

Regarding the SOC states $\ket{J_z=\pm1,0}$ as ground states of a single $\mathrm{Fe}^{2+}$ ion implies that the Jahn-Teller (JT) effect has been ignored. This approximation is justified for $t_{2g}$ orbitals, which have a relatively weak JT effect. Concretely, the magnitude of JT distortions scales with the orbital-ligand overlap. In the octahedral environment, $e_g$ orbitals (directed toward ligands) undergo strong distortion when unevenly occupied,
whereas $t_{2g}$ orbitals (oriented between ligands) display substantially weaker JT effects. 
Moreover, when SOC is dominant over the JT effect, the SOC splits the degenerate $t_{2g}$ manifold into well-separated spin-orbit states described by spin-orbit entangled ``pseudospins'', 
and the lattice remains largely undistorted.
The SOC can not only suppress the JT distortion but also enable effectively larger spin-orbital pseudo-spins \cite{witczak-krempa2014CorrelatedQuantumPhenomena}.
Therefore, we could safely neglect JT effects in our analysis up to the leading order. 
Notably, quantitative analysis of SOC versus JT competition reveals 
that JT effects are generally suppressed for typical SOC strengths in 4d and 5d ions \cite{minarro2025EmergentOrbitalDynamics}.

\subsection{Magnetic Moments}
To reveal the effect of the quadrupole moment, 
we construct a dipole-order-free basis in the $J=1$ manifold as,
\begin{equation}
\begin{split}
& \ket{x}=\frac{1}{\sqrt{2}}(\ket{J_z=-1}-\ket{J_z=+1}), \\
& \ket{y}=\frac{i}{\sqrt{2}}(\ket{J_z=-1}+\ket{J_z=+1}), \\
& \ket{z}=\ket{J_z=0},
\end{split}
\end{equation}
with a general state in $J=1$ manifold expressed by 
\begin{equation}
\ket{\psi}=b_x\ket{x}+b_y\ket{y}+b_z\ket{z},
\label{eq:psi}
\end{equation}
where the coefficient vector ${\bm{b}=(b_x,b_y,b_z)}$ is in general complex.
Since the total angular momentum of ${J = 1}$ allows for the presence of 
the high-rank magnetic moments, 
it can accommodate both the dipole and quadrupole moments. 
The dipole moment is directly related to the local moment $\bm{J}$ itself,
while the quadrupole moment $Q_{\mu\nu}$ is given by the rank-2 tensor,
\begin{equation}
    Q_{\mu\nu}=\frac{1}{2}\{J_\mu,J_\nu\}-\frac{\bm{J}^2}{3}\delta_{\mu\nu},\quad \mu,\nu\in\{x,y,z\},
\end{equation}
where $\delta_{\mu\nu}$ is the Kronecker delta symbol. 
$Q_{\mu\nu}$ can be understood as the 2nd-order Steven's operator, which reveals the underlying spin quadrupole orders in the system \cite{kusunose2008DescriptionMultipole}.
In the state $\ket{\psi}$, the dipole and quadrupole orders 
are evaluated as, 
\begin{equation}
    \braket{\psi|\bm{J}|\psi}=-i\bm{b}^*\times\bm{b},
    \label{eq:mag-dipole}
\end{equation}
and
\begin{equation}
    \braket{\psi|Q_{\mu\nu}|\psi}=\frac{1}{3}\delta_{\mu\nu}-\frac{1}{2}(b^*_\mu b_\nu+b^*_\nu b_\mu),
    \label{eq:mag-quadrupole}
\end{equation}
indicating that their existence and disappearance can be controlled by tuning the vector $\bm b$. 
The symmetric tensor $Q_{\mu\nu}$ has five independent components, 
$Q_{3z^2-r^2}$, $Q_{x^2-y^2}$, $Q_{xy}$, $Q_{yz}$, and $Q_{zx}$. 
The presence of the quadrupole moments in the local Hilbert space 
suggests that these high-spin systems could exhibit richer ferroelectric 
behaviors beyond those explained by the inverse Dzyaloshinskii-Moriya mechanism. 
Moreover, when the dipole moment vanishes under the condition $ \bm{b}^* \parallel \bm{b} $, 
the state $ |\psi\rangle $ only retains the quadrupole order. 
This enables the exploration of the mechanism from which
the ferroelectricity arises purely from the quadrupole orders.

To incorporate both the dipolar order and quadrupolar order for the Mott insulator, 
we treat these orders as a mean field coupled to the local spin $\bm{J}$ in the relevant channel.
We consider a generic local coupling, $H_0=J_1\hat{\bm{e}}\cdot\bm{J}+J_2(\hat{\bm{e}}\cdot\bm{J})^2$,
where the magnetic ordering vector $\hat{\bm{e}}=(\sin \theta \cos \phi,\sin \theta \sin \phi,\cos \theta)$ 
for the ${J=1}$ local momentum and the couplings ${J_{1,2}>0}$.
Under the application of $H_0$, the ground state of the $d^6$ electrons of 
the Fe$^{2+}$ ion can be described as $\ket{\psi}$ with a $J_{1,2}$-dependent vector 
$\bm b$, (see App.~\ref{app:single site Ham}).
In particular, when ${J_1<J_2}$, the vector $\bm{b}$ is identical to the unit vector 
$\hat{\bm{e}}$, and a spin state without the magnetic dipole orders is established.

\begin{figure}
    \centering
    \includegraphics[width=0.95\linewidth]{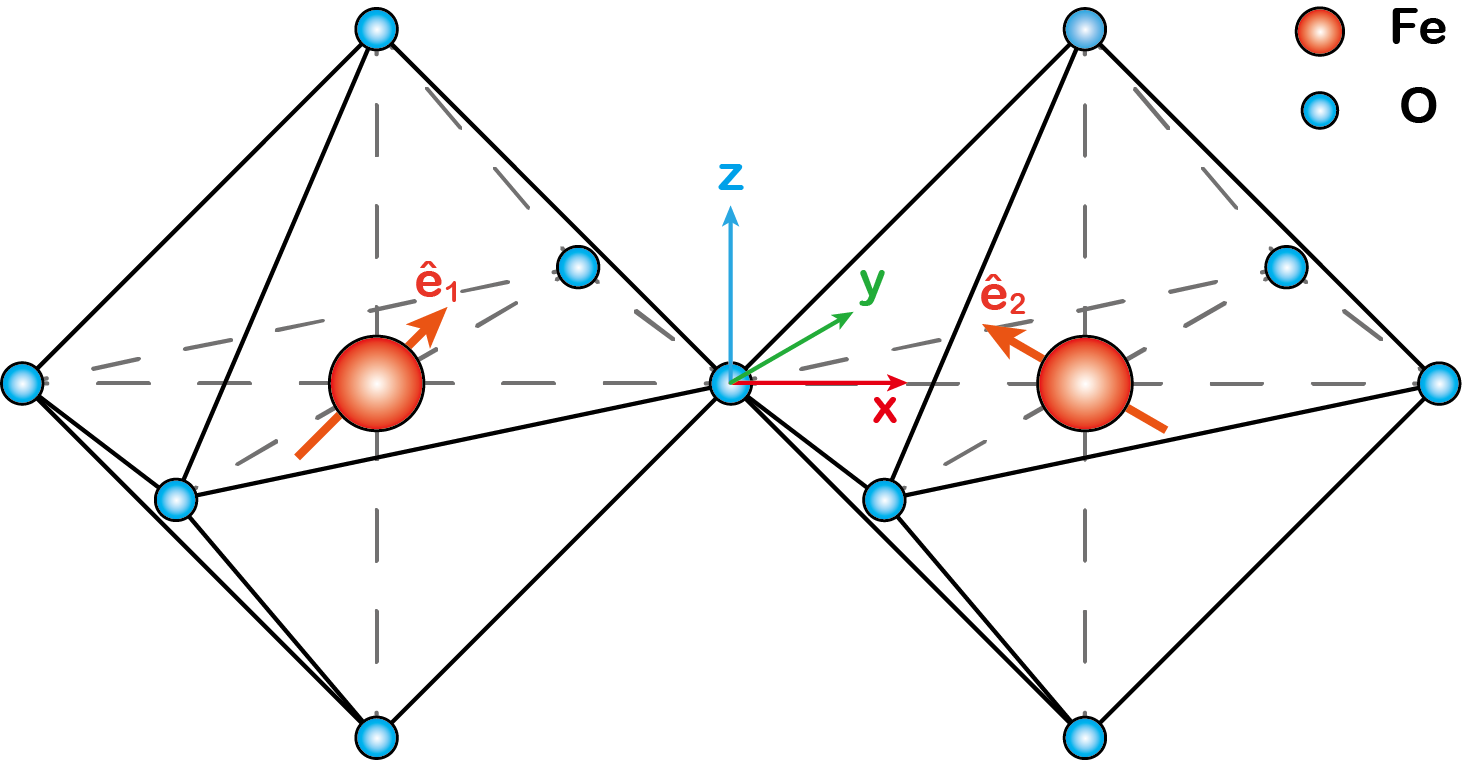}
    \caption{The corner-sharing octahedral cluster. 
    The red solid arrow at each Fe ion indicates the direction of non-collinear magnetic ordering $\hat{\bm{e}}$. 
    The vector determines the ``direction'' of the quadrupolar order when the dipolar order is quenched.}
    \label{fig:two-octahedron}
\end{figure}

\section{Three-site cluster}
\label{sec:Three-site cluster}
We consider a minimal model consisting of 
two corner-sharing octahedra (see Fig.~\ref{fig:two-octahedron}). 
Without loss of generality, the cluster is oriented along the $x$-direction, 
and the orbital quantization is along the $z$-direction. 
Due to the finite intersite overlap between the $3d$ orbitals of 
the Fe sites and the $2p$ orbitals of the oxygen (O) site, i.e.  
the $\pi$ bonding from the $t_{2g}$ orbitals  
and the $\sigma$ bonding from the $e_g$ orbital, the electron hopping is described by the following perturbative Hamiltonian,
\begin{equation}
\begin{aligned}
&H_\text{hop}=\sum_{\sigma;\alpha=y,z} t
\left(
d_{1,x\alpha,\sigma}^\dagger p_{\alpha\sigma}^{}-
d_{2,x\alpha,\sigma}^\dagger p_{\alpha\sigma}^{}
\right)
\\& \quad
+ \sum_{\sigma} t_0 \left(
d_{1,3x^2,\sigma}^\dagger p_{x\sigma}^{}
-
d_{2,3x^2,\sigma}^\dagger p_{x\sigma}^{}
\right) +\text{h.c.},
\end{aligned}
\label{eq:Hhop}
\end{equation}
where $t$ and $t_0$ are the hopping parameters, $d_{i,a,\sigma}^\dagger$ 
($i=1,2$, $a=xy,xz,3x^2-r^2$, $\sigma=\uparrow,\downarrow$) 
creates a hole with the spin $\sigma$ in the $d$-orbital of the $i$-th Fe site, 
 $p_{\beta\sigma}^\dagger$ ($\beta=x,y,z$) creates a hole 
 with the spin $\sigma$ in the $p$-orbital of O, 
 and $3x^2$ refers to the $3x^2-r^2$ orbital.    
The $3x^2$ orbital is related to the $3z^2$ and $x^2-y^2$ orbitals 
with the relation ${d^\dagger_{i,3x^2,\sigma}=\frac{\sqrt{3}}{2}d^\dagger_{i,x^2-y^2,\sigma}
- \frac{1}{2} d^\dagger_{i,3z^2 ,\sigma}}$. 
Therefore, the total Hamiltonian for the three-site cluster is 
\begin{equation}
H=H_{\text{Fe},1}+H_{\text{Fe},2}+H_{\text{O}}+H_{\text{hop}} , 
\end{equation}
where the first three terms describe the onsite electron correlations 
leading to different ground states, such as $\ket{\psi}$ in Eq.~\eqref{eq:psi} 
for the $d^6$ electrons at the Fe site.

We take the electron-doped Fe-based Mott insulator 
as an example, where the $d^6$ and $d^7$ electron configurations 
are considered for the two Fe sites, and the net charge order is assumed to be absent. 
For the perturbation of the $ d^6 $ ($ d^7 $) configuration, 
we obtain an intermediate configuration of $ d^7 $ ($ d^8 $) 
due to the virtual hopping process. Furthermore, it is ready to notice that, 
owing to the strong onsite interaction and SOC, the electron configurations 
have to be reconstructed, whose details are given in App.~\ref{app:spin configuration}.

In the hole representation, the ground state of the four holes ($d^6$) 
is given by $\ket{\psi}$ in Eq.~\eqref{eq:psi}. 
The low-energy physics of three holes ($d^7$) can be effectively described 
by a total angular momentum ${ J = 1/2 }$. 
We denote $ \ket{J_z = 1/2} \equiv \ket{\Uparrow} $ and 
$ \ket{J_z = -1/2} \equiv \ket{\Downarrow} $, 
and the ground state is $ \ket{\phi_i} = a_{i\Uparrow} \ket{\Uparrow} + a_{i\Downarrow} \ket{\Downarrow} $, 
where $ \bm{a}_i = (a_{i\Uparrow}, a_{i\Downarrow}) $ is a normalized complex-valued vector. 
Hence, the ground state for the cluster is given by
$\ket{\Psi_0}=\frac{1}{\sqrt{2}}\left(\ket{\phi_1,\psi_2}+\ket{\psi_1,\phi_2}\right)$.
The notation $\ket{\phi_1,\psi_2}$ is a short-handed notation of 
$\mathscr{P}\ket{\phi_1}\otimes\ket{0}\otimes\ket{\psi_2}$, 
where $\mathscr{P}$ is an operator making sure the fermionic anti-symmetry. 
We have assumed the oxygen site is filled with electrons, i.e. absent of holes.

Due to $H_{\text{hop}}$, an electron on the Fe sites can hop to the oxygen site, 
or equivalently, a hole on the Fe site can hop to the oxygen site. With this hybridization process, 
the first-order perturbed (unnormalized) state is given by 
\begin{equation}\label{eq:perturbed_state}
\begin{aligned}
\ket{\Psi}&=\ket{\phi_1,\psi_2} + \ket{\psi_1,\phi_2}
\\&
+\frac{1}{\Delta}
\sum_{d^8,\alpha}
\ket{d^8_1,p_\alpha,\psi_2}\braket{d^8_1,p_\alpha,\psi_2|H_{\text{hop}}|\phi_1,\psi_2} 
\\&
+\frac{1}{\Delta}
\sum_{d^7,\alpha}\ket{\phi_1,p_\alpha,d^7_2}
\braket{\phi_1,p_\alpha,d^7_2|H_{\text{hop}}|\phi_1,\psi_2} 
\\&
+\frac{1}{\Delta}
\sum_{d^8,\alpha}\ket{\psi_1,p_\alpha,d^8_2}
\braket{\psi_1,p_\alpha,d^8_2|H_{\text{hop}}|\psi_1,\phi_2} \\
&+\frac{1}{\Delta}
\sum_{d^7,\alpha}\ket{d^7_1,p_\alpha,\phi_2}
\braket{d^7_1,p_\alpha,\phi_2|H_{\text{hop}}|\psi_1,\phi_2},
\end{aligned}
\end{equation}
where the summation runs over all possible 2-hole ($d^8$) or 3-hole ($d^7$) states, 
and all $2p$ orbitals of the oxygen. The energy separation between a $2p$ orbital 
and $d$-orbitals is approximated as $\Delta$.
 
\section{Electric polarization}
\label{sec:Electric polarization}
For the perturbed state $\ket{\Psi}$, 
the electric polarization $\bm{P}=\braket{\Psi|e\bm{r}|\Psi}$. 
The operator $\bm{r}$ is a many-body operator, which is a sum of the position 
vectors of all particles $\bm{r}=\sum_{n}\bm{r}_n$ if 
using the first quantization language. After a detailed analysis, 
we find out six non-vanishing contributions. The first four contributions are from 
\begin{equation}
\begin{split}
&\braket{\phi_1,\psi_2|\bm{r}|d^8_1,p_\alpha,\psi_2}=\braket{\phi_1|\bm{r}|d^8_1,p_\alpha},\\
&\braket{\phi_1,\psi_2|\bm{r}|\phi_1,p_\alpha,d^7_2}=\braket{\psi_2|\bm{r}|p_\alpha,d^7_2},\\
&\braket{\psi_1,\phi_2|\bm{r}|\psi_1,p_\alpha,d^8_2}=\braket{\phi_2|\bm{r}|p_\alpha,d^8_2},\\
&\braket{\psi_1,\phi_2|\bm{r}|d^7_1,p_\alpha,\phi_2}=\braket{\psi_1|\bm{r}|d^7_1,p_\alpha},
\end{split}
\end{equation}
each depending only on the single Fe-site parameters. 
Hence, we refer to these as the onsite contributions to the electric polarization 
and denote them as $\bm{P}_{\text{on}}$. The remaining terms, referred to as 
the inter-site contributions and denoted as $\bm{P}_{\text{int}}$, 
are given by
\begin{equation}
\begin{split}
&\braket{\phi_1,\psi_2|\bm{r}|d^7_1,p_\alpha,\phi_2},\\
&\braket{\psi_1,\phi_2|\bm{r}|\phi_1,p_\alpha,d^7_2},
\end{split}
\end{equation}
which have a hybrid form involving $\bm{b}$ and $\bm{a}$.
Other terms in $\braket{\Psi|\bm{r}|\Psi}$ vanish due to the orthogonality 
and the parity requirements. 

The onsite contributions $\bm{P}_{\text{on}}$ arise purely 
from the hybridization between the Fe ions and the ligand O ion. 
Since the onsite contributions are obtained for each Fe site, 
we can discuss their contribution separately, i.e., 
$\bm{P}_{\text{on}}=\bm{P}_{1,\text{on}}-\bm{P}_{2,\text{on}}$. 
The onsite contribution from the hybridization between 
the $i$-th Fe atom and the O atom can be obtained from 
the perturbative calculation 
\begin{eqnarray}
\label{eq:onsite}
\tilde{P}_{i,\text{on}}^x&=&
\frac{49}{\sqrt{3}}\braket{Q_{3z^2-r^2}}_i +
\braket{Q_{x^2-y^2}}_i  -3(\sqrt{3}-1){t_0}{t}^{-1}
\nonumber \\ 
&& \times [2\braket{Q_{xy}}_i
+\braket{J_x}_i^{(d^6)}+\braket{J_y}_i^{(d^6)} ], \\
\tilde{P}_{i,\text{on}}^y&=&-\braket{Q_{xy}}_i, \\ 
\tilde{P}_{i,\text{on}}^z&=&\braket{J_x}_i^{(d^6)}-8\braket{J_x}_i^{(d^7)},
\end{eqnarray}
where we have introduced $\tilde{P}^\mu_{i,\text{on}}=P^\mu_{i,\text{on}}/s_\mu$ with $s_x=4c$, $s_y=2c$, $s_z=10c$, and ${c=e t I/(480 \Delta)}$ accounting for the anisotropy. 
The integral $I$ is the overlap between $d$ orbitals and $p$ orbitals and is given by 
\begin{equation}
    I=\int\diff^3\bm{r}\,d_{xy}(\bm{r})xp_y(\bm{r}),
\end{equation}
and its cyclic permutations of $x,y,z$. It is clear that both the quadrupole and dipole of 
the ${J=1}$ states and the dipole of the ${J=1/2}$ states have contributions to $\bm{P}_{\text{on}}$. 
This result indicates that $\bm{P}_{\text{on}}$ is finite as long as 
$\bm{P}_{1,\text{on}}\neq\bm{P}_{2,\text{on}}$, 
corresponding to the nonuniform alignment of the quadrupoles and dipole moments. 
The underlying origin is that the finite $\bm P_{\text{on}}$ arises 
from the inversion symmetry breaking of the local moments 
via the non-uniform ordering, instead of directly from an imbalanced charge density distribution. 
Thus, it belongs to the so-called improper ferroelectricity. 
Moreover, these onsite contributions to the electric polarization, 
which is absent in the case considered in the original work in 
Ref.~\cite{katsuraSpinCurrentMagnetoelectric2005}, 
are essential in the many-electron configurations. 
Moreover, we can make a zeroth-order estimation to the order of magnitude for $P^{x,y,z}$. 
With the lattice constant $a= 5 $\AA, 
$\bm{P} \sim c \sim 10^{-5} (t/\Delta)\ C/m^2$, 
which is consistent with the results of
$\rm Ga_{2-x}Fe_xO_3$ \cite{popov1998MagnetoelectricEffect}.

For the inter-site contribution $\bm{P}_{\text{int}}$, two Fe ions are hybridized. 
The full expressions of $\bm{P}_{\text{int}}$ in terms of $\bm{a}_i$ and $\bm{b}_i$ 
($i=1,2$) are listed in App.~\ref{app:Full P}. 
Here, to explain the essence of the multipolar ferroelectricity, 
we simplify our discussion to two different cases: 
(i) $d^7$ states are uniform; 
(ii) $d^6$ states are uniform.

In case (i), we set ${\bm{a}_1=\bm{a}_2=(1,0)}$, such that the dipolar order of 
the $d^7$ states are collinear and along the $z$-direction. 
The inter-site contribution $\bm{P}_{\text{int}}$ can be reduced as follows
\begin{align}
    \tilde{P}^x_{\text{int}} & = 0, \\
     \tilde{P}^y_{\text{int}} & =-\left[\hat{\bm{x}}\times\left(\bm{b}_1^*\times\bm{b}_2^{} \right)\right]_y
      -2ib_{1z}^*b_{2z}+\text{h.c.}, \\
     \tilde{P}^z_{\text{int}} & = -\left[\hat{\bm{x}}\times\left(\bm{b}_1^*\times\bm{b}_2^{} \right)\right]_z+i(b_{1y}^*b_{2z}^{}+b_{1z}^*b_{2y}^{})+\text{h.c.}.
\end{align}
The first terms of $\tilde{P}^y_{\text{int}}$ and $\tilde{P}^y_{\text{int}}$ 
resemble the inverse Dzyaloshinskii-Moriya mechanism~\cite{katsuraSpinCurrentMagnetoelectric2005}, 
while the second terms endow the mechanism with some additional effects. 
For a generic $\bm{b}_1$ and $\bm{b}_2$, the dipolar and quadrupolar orders are concomitant. 
We single out the quadrupolar order by choosing $\bm{b}_1=\hat{\bm{e}}_1$     
and $\bm{b}_2=\hat{\bm{e}}_2$ as the real-valued unit vectors, 
in which the dipolar orders are quenched. 
We then establish the multipolar version of the inverse 
Dzyaloshinskii-Moriya mechanism,
\begin{equation}
    \tilde{\bm{P}}_{\text{int}}\sim \hat{\bm{x}}\times\left(\hat{\bm{e}}_1\times\hat{\bm{e}}_2\right).
\end{equation}
This indicates that in a system without the non-collinear magnetic orders, 
a finite electric polarization can still be generated from the non-uniform quadrupolar orders. 
This clearly goes beyond the scope of the original inverse Dzyaloshinskii-Moriya mechanism~\cite{katsuraSpinCurrentMagnetoelectric2005}. 
The difference is that our mechanism is now determined by the underlying vector 
$\hat{\bm{e}}$ for the quadrupole moment $Q_{\mu\nu}$ 
rather than the direction of the dipole moment.

In case (ii), the $d^6$ states are taken to be uniform. 
We discuss how the quadrupolar orders and the dipolar orders 
of the $d^6$ states modify the inverse Dzyaloshinskii-Moriya mechanism 
of the $d^7$ states. 
Fixing ${\bm{b}_1=\bm{b}_2=(0,0,1)}$, 
we obtain $\tilde{P}^x_{\text{int}}=\tilde{P}^z_{\text{int}}=0$, 
and 
\begin{equation}
    \tilde{P}^y_{\text{int}}=-i\left(a_{1\Uparrow}^*a_{2\Uparrow}
    -a_{1\Uparrow}a_{2\Uparrow}^*-a_{1\Downarrow}^*a_{2\Downarrow}
    +a_{1\Downarrow}a_{2\Downarrow}^*\right).
\end{equation}
A finite electric polarization is generated by the non-collinear order 
of the ${J=1/2}$ dipole moments of the $d^7$ states. 
We find that 
\begin{equation}
\tilde{P}^y_{\text{int}}\sim \sin\theta_1 \sin\theta_2 \sin(\phi_1-\phi_2),
\end{equation}
with the two spinors of the $d^7$ states as ${\bm{a}_1=}\frac{\alpha}{|\alpha|}
e^{-i\phi}(\sin\frac{\theta_1}{2},e^{i\phi_1}\cos\frac{\theta_1}{2})$,
$\bm{a}_2=(\sin\frac{\theta_2}{2}$, $e^{i\phi_2}\cos\frac{\theta_2}{2})$, 
where ${\phi=\phi_1-\phi_2}$ 
and $\alpha=e^{-i\phi/2}\cos\frac{\theta_1}{2}\cos\frac{\theta_2}{2}
+e^{i\phi/2}\sin\frac{\theta_1}{2}\sin\frac{\theta_2}{2}$.
This expression of $\tilde{P}^y_{\text{int}}$ is consistent with the
 earlier results in Ref.~\cite{katsuraSpinCurrentMagnetoelectric2005}.
It demonstrates the crossover from the multipolar origin to the original 
inverse Dzyaloshinskii-Moriya mechanism for the ferroelectricity.

\section{Discussion and Conclusion}
\label{sec:Discussion}
In this work, we consider a minimal cluster to 
demonstrate that quadrupolar-ordered magnets can make  
a pronounced contribution to ferroelectricity. 
The original inverse Dzyaloshinskii-Moriya mechanism reveals the presence of the 
finite electric polarization under the non-collinear magnetic dipolar orders. 
We have generalized this formalism to the cases with higher-rank magnetic 
moments. By tuning the quadrupolar order together with the dipolar order, 
we obtain a crossover or transition from the multipolar ferroelectricity 
to the dipolar ferroelectricity. Moreover, the coexistence of the quadrupolar 
and dipolar orders significantly modifies the inverse Dzyaloshinskii-Moriya mechanism. 
In general, we expect the multipolar ferroelectricity to occur widely in 
the spin-orbit-coupled Mott insulators with large $J$ moments. 
These include
the $4d/5d$ materials \cite{Voleti2020Multipolar,Pradhan2024Multipolar,chenExoticPhasesInduced2010,chenSpinorbitCoupling$d^2$2011,yamadaEmergent$mathrmSU4$Symmetry2018,caoFrontiers4D5Dtransition2013}, the $4f/5f$ magnets \cite{Leonid2021Hidden,Santini2009Multipolar,kusunose2008DescriptionMultipole}, some $3d$ transition metal compounds\cite{popov1998MagnetoelectricEffect,popov1999MagneticfieldinducedToroidal}, and hybrid orbital systems \cite{hayami2018MicroscopicDescription,kuramoto2009MultipoleOrders,hayami2018ClassificationAtomicscale}.

The interplay between the high-rank multipolar orders and the dipolar order 
can lead to rather rich behaviors at zero and finite temperatures. The 
quadrupolar order often does not break time reversal, while the dipolar 
order breaks time reversal. From the Ginzburg-Landau theory \cite{chenExoticPhasesInduced2010}, the dipolar order could induce the quadrupolar order
as a subsidiary order, but the reverse is not true. Moreover, the quadrupolar 
order could persist up to a higher temperature than the dipolar order \cite{kim2024QuantumSpin,hirai2020DetectionMultipolar}. 
In terms of the ferroelectricity, the quadrupolar ferroelectricity 
could occur at a higher temperature than the dipolar ferroelectricity. 
The quadrupolar ferroelectricity further enriches the magnetoelectric response. 
The previously proposed ``double-leaf Riemann surface topological converse magnetoelectricity''
for the dipolar ferroelectricity \cite{zhou2024DoubleleafRiemann}, which dictates the half-periodic response of the 
magnetic dipolar moment to the external electric field, could generalize to the quadrupolar ferroelectricity. More interestingly, in the case when the quadrupolar order itself is induced by the dipolar order, such a double-leaf Riemann surface topological 
converse magnetoelectricity could be further complicated because the response of 
dipolar order to the variation of the quadrupolar order is another level of double-leaf Riemann surface. 
Likewise, the previously proposed
topological Roman surface by dipolar-order-induced ferroelectric polarization \cite{liu2022PhysicalRealization}
could further be complicated for the quadrupolar ferroelectricity 
with the intertwined structures of the dipolar and quadrupolar orders.

To summarize, our work exhibits a multipolar origin for the ferroelectricity in Mott insulators, 
even in the absence of the dipolar order. 
It provides a framework that unifies dipolar and multipolar-based electric polarization. 
This framework may provide a better understanding of unconventional 
ferroelectric materials and the experimental guidance for the design of materials 
with tunable multiferroic properties.

\section*{Acknowledgments}
This work is supported by the MOST of China with 
Grant No.~2021YFA1400300, and by the Fundamental Research Funds for the Central Universities, 
Peking University. 

\clearpage
\onecolumngrid
\appendix
\section{Review of atomic orbitals}
In this section, we give a brief review of atomic orbitals and introduce our notations. For those who are familiar with atomic orbitals, this section can be skipped. 

When we talk about atomic orbitals, we are talking about the eigenstates of hydrogen and hydrogen-like atoms. Basic quantum mechanics tells us that real space wave functions of atomic orbitals are given by $\psi_{n,l,m}(\bm{r})=R_{n,l}(r)Y_{l,m}(\theta,\phi)$. The angular part $Y_{l,m}(\theta,\phi)$ is the complex-valued spherical harmonics, while the radial part $R_{n,l}(r)$ is real-valued. Eigenstates in the form of $\psi_{n,l,m}(\bm{r})$ are called complex orbitals. In practice, people usually use real orbitals, which are linear combinations of complex orbitals. Following the Condon–Shortley phase convention, we define real orbitals as
\begin{equation}
    \psi^{\text{real}}_{n,l,m}(\bm{r})=
    \begin{cases}
        \sqrt{2}(-1)^{m}\Re\psi_{n,l,|m|}(\bm{r})=\frac{1}{\sqrt{2}}\left[\psi_{n,l,-|m|}(\bm{r})+(-1)^{m}\psi_{n,l,|m|}(\bm{r})\right], &\quad m>0 \\
        \psi_{n,l,|m|}(\bm{r}),&\quad m=0 \\
        \sqrt{2}(-1)^{m}\Im\psi_{n,l,|m|}(\bm{r})=\frac{i}{\sqrt{2}}\left[\psi_{n,l,-|m|}(\bm{r})-(-1)^{m}\psi_{n,l,|m|}(\bm{r})\right],&\quad m<0
    \end{cases}
\end{equation}
These orbitals are purely real and usually labeled by harmonic polynomials. For example, $p$ orbitals ($l=1$) are given by
\begin{equation}
\begin{split}
& \psi_{n,x}=\psi^{\text{real}}_{n,1,+1}=\frac{1}{\sqrt{2}}(\psi_{n,1,-1}-\psi_{n,1,+1})=R_{n,1}\sqrt{\frac{3}{4\pi}}\frac{x}{r}, \\
& \psi_{n,z}=\psi^{\text{real}}_{n,1,0}=\psi_{n,1,0}=R_{n,1}\sqrt{\frac{3}{4\pi}}\frac{z}{r}, \\
& \psi_{n,y}=\psi^{\text{real}}_{n,1,-1}=\frac{i}{\sqrt{2}}(\psi_{n,l,-1}+\psi_{n,l,+1})=R_{n,1}\sqrt{\frac{3}{4\pi}}\frac{y}{r}.
\end{split}
\end{equation}
$d$ orbitals ($l=2$) are given by
\begin{equation}
\begin{split}
& \psi_{n,x^2-y^2}=\psi_{n,2,+2}^{\text{real}}=\frac{1}{\sqrt{2}}(\psi_{n,2,-2}+\psi_{n,2,+2})=R_{n,2}\frac{1}{4}\sqrt{\frac{15}{\pi}}\frac{x^2-y^2}{r^2}, \\
& \psi_{n,zx}=\psi_{n,2,+1}^{\text{real}}=\frac{1}{\sqrt{2}}(\psi_{n,2,-1}-\psi_{n,2,+1})=R_{n,2}\frac{1}{2}\sqrt{\frac{15}{\pi}}\frac{zx}{r^2}, \\
& \psi_{n,3z^2-r^2}=\psi_{n,2,0}^{\text{real}}=\psi_{n,2,0}=R_{n,2}\frac{1}{2}\sqrt{\frac{5}{\pi}}\frac{3z^2-r^2}{r^2}, \\
& \psi_{n,yz}=\psi_{n,2,-1}^{\text{real}}=\frac{i}{\sqrt{2}}(\psi_{n,2,-1}+\psi_{n,2,+1})=R_{n,2}\frac{1}{2}\sqrt{\frac{15}{\pi}}\frac{yz}{r^2}, \\
& \psi_{n,xy}=\psi_{n,2,-2}^{\text{real}}=\frac{i}{\sqrt{2}}(\psi_{n,2,-2}-\psi_{n,2,+2})=R_{n,2}\frac{1}{2}\sqrt{\frac{15}{\pi}}\frac{xy}{r^2}.
\end{split}
\end{equation}

\section{Many-body states for different electron configurations}
\label{app:spin configuration}
Throughout our paper, we are working on the $3d^n$ configurations of Fe atoms. In this environment of an octahedral crystal field, the crystal field splits the fivefold degenerate $d$-orbitals into a $t_{2g}$ triplet as the ground state and an $e_g$ doublet, separated by an energy gap $\Delta_{\text{cry}}$. We consider the situation in which $\Delta_{\text{cry}}$ is smaller than the onsite repulsion of each orbital. Then electrons tend to occupy different orbitals rather than fill the $t_{2g}$. In this circumstance, we analyze the many-body states for $d^6$, $d^7$, and $d^8$ electron configurations.

\subsection{\texorpdfstring{$d^6=t_{2g}^4e_{g}^2$}{} configuration}
For $d^6$ electron configuration, there are 2 electrons in $e_{g}$ orbitals and 4 electrons in $t_{2g}$ orbitals. Thus, we can denote $d^6$ as $t_{2g}^4e_{g}^2$. There are 3 possibilities to arrange the 4 electrons in $t_{2g}$ orbitals, giving rise to threefold orbital states. Since 6 electrons are more than half-filling, it is more convenient to use the hole representation. In the hole representation, threefold orbital states are given by
\begin{equation}\label{eq:ABC}
\begin{split}
    & \ket{a}=A^\dagger|0\rangle=d^\dagger_{3z^2-r^2}d^\dagger_{x^2-y^2}d_{zx}^\dagger d_{xy}^\dagger\ket{0}, 
    \\
    & \ket{b}=B^\dagger|0\rangle=d^\dagger_{3z^2-r^2}d^\dagger_{x^2-y^2}d_{xy}^\dagger d_{yz}^\dagger\ket{0}, 
    \\
    & \ket{c}=C^\dagger|0\rangle=d^\dagger_{3z^2-r^2}d^\dagger_{x^2-y^2}d_{yz}^\dagger d_{zx}^\dagger\ket{0}, 
\end{split}
\end{equation}
where $d_{\alpha}^\dagger$ creates a hole at the orbital $\alpha$ with arbitrary spin and $\ket{0}$ is the vacuum state of holes. The projection of the physical orbital angular momentum $\bm{L}_{\text{physical}}$ on the threefold orbital states gives an effective angular momentum $\bm{L}\cong-\bm{L}_{\text{physical}}$ with the quantum number $L=1$. The eigenstates of $L_z$ are
\begin{equation}\label{eq:L=1}
\begin{split}
& \ket{L_z=+1}=\frac{1}{\sqrt{2}}(\ket{a}+i\ket{b}), \\
& \ket{L_z=0}=\ket{c}, \\
& \ket{L_z=-1}=\frac{1}{\sqrt{2}}(\ket{a}-i\ket{b}).
\end{split}
\end{equation}
The first Hund's rule tells us the $d^6$ configuration has a total spin $\bm{S}$ with the spin quantum number $S=2$. The eigenstates of $S_z$ are given by
\begin{equation}\label{eq:S=2}
\begin{split}
& \ket{S_z=2}=\ket{\uparrow\uparrow\uparrow\uparrow}, \\
& \ket{S_z=1}=\frac{1}{2}(\ket{\uparrow\uparrow\uparrow\downarrow}+\ket{\uparrow\uparrow\downarrow\uparrow}+\ket{\uparrow\downarrow\uparrow\uparrow}+\ket{\downarrow\uparrow\uparrow\uparrow}), \\
& \ket{S_z=0}=\frac{1}{\sqrt{6}}\left(\ket{\uparrow\uparrow\downarrow\downarrow}+\ket{\uparrow\downarrow\uparrow\downarrow}+\ket{\uparrow\downarrow\downarrow\uparrow}+\ket{\downarrow\uparrow\uparrow\downarrow}+\ket{\downarrow\uparrow\downarrow\uparrow}+\ket{\downarrow\downarrow\uparrow\uparrow}\right), \\
& \ket{S_z=-1}=\frac{1}{2}(\ket{\downarrow\uparrow\uparrow\uparrow}+\ket{\uparrow\downarrow\uparrow\uparrow}+\ket{\uparrow\uparrow\downarrow\uparrow}+\ket{\uparrow\uparrow\uparrow\downarrow}), \\
& \ket{S_z=-2}=\ket{\downarrow\downarrow\downarrow\downarrow}.
\end{split}
\end{equation}
Then, we consider the spin-orbital coupling (SOC) $H_{\text{SOC}}=\lambda\bm{L}\cdot\bm{S}$ with a coupling constant $\lambda>0$. We can define the total angular momentum $\bm{J}=\bm{L}+\bm{S}$ and write $H_{\text{SOC}}=\lambda(\bm{J}^2-\bm{L}^2-\bm{S}^2)/2$. Thus, the SOC ground states are given by the quantum number $J=1$. Utilizing the Clebsh-Gordan coefficients, we obtain
\begin{equation}
\begin{split}
& \ket{J_z=1}=\sqrt{\frac{1}{10}}\ket{L_z=1,S_z=0}-\sqrt{\frac{3}{10}}\ket{L_z=0,S_z=1}+\sqrt{\frac{3}{5}}\ket{L_z=-1,S_z=2}, \\
& \ket{J_z=0}=\sqrt{\frac{3}{10}}\ket{L_z=1,S_z=-1}-\sqrt{\frac{2}{5}}\ket{L_z=0,S_z=0}+\sqrt{\frac{3}{10}}\ket{L_z=-1,S_z=1}, \\
& \ket{J_z=-1}=\sqrt{\frac{3}{5}}\ket{L_z=1,S_z=-2}-\sqrt{\frac{3}{10}}\ket{L_z=0,S_z=-1}+\sqrt{\frac{1}{10}}\ket{L_z=-1,S_z=0}.
\end{split}
\end{equation}
We also define three real-valued states from the threefold $J=1$ states,
\begin{equation}\label{eq:XYZ}
\begin{split}
& \ket{x}=X^\dagger\ket{0}=\frac{1}{\sqrt{2}}(\ket{J_z=-1}-\ket{J_z=+1}), \\
& \ket{y}=Y^\dagger\ket{0}=\frac{i}{\sqrt{2}}(\ket{J_z=-1}+\ket{J_z=+1}), \\
& \ket{z}=Z^\dagger\ket{0}=\ket{J_z=0},
\end{split}
\end{equation}
Combining \cref{eq:ABC,eq:L=1,eq:S=2}, we can express the \cref{eq:XYZ} states in terms of creation operators of hole,
\begin{equation}
\begin{aligned}
\hat{X}&=\frac{1}{4 \sqrt{15}}{d^\dagger_{z^2,\uparrow} d^\dagger_{x^2-y^2,\uparrow} \left(-3 d^\dagger_{yz,\uparrow} d^\dagger_{zx,\downarrow}-3 d^\dagger_{yz,\downarrow} d^\dagger_{zx,\uparrow}+2 i d^\dagger_{xy,\downarrow} d^\dagger_{yz,\downarrow}-6 i d^\dagger_{xy,\uparrow} d^\dagger_{yz,\uparrow}+6 d^\dagger_{zx,\uparrow} d^\dagger_{xy,\uparrow}\right)}\\
&+\frac{1}{4 \sqrt{15}}{d^\dagger_{z^2,\downarrow} d^\dagger_{x^2-y^2,\downarrow} \left(3 d^\dagger_{yz,\uparrow} d^\dagger_{zx,\downarrow}+3 d^\dagger_{yz,\downarrow} d^\dagger_{zx,\uparrow}-6 i d^\dagger_{xy,\downarrow} d^\dagger_{yz,\downarrow}-6 d^\dagger_{zx,\downarrow} d^\dagger_{xy,\downarrow}+2 i d^\dagger_{xy,\uparrow} d^\dagger_{yz,\uparrow}\right)}\\
&+\frac{1}{4 \sqrt{15}}{d^\dagger_{z^2,\uparrow} d^\dagger_{x^2-y^2,\downarrow} \left(2 i d^\dagger_{xy,\uparrow} d^\dagger_{yz,\downarrow}+2 i d^\dagger_{xy,\downarrow} d^\dagger_{yz,\uparrow}+3 d^\dagger_{yz,\downarrow} d^\dagger_{zx,\downarrow}-3 d^\dagger_{yz,\uparrow} d^\dagger_{zx,\uparrow}\right)}\\
&+\frac{1}{4 \sqrt{15}}{d^\dagger_{z^2,\downarrow} d^\dagger_{x^2-y^2,\uparrow} \left(2 i d^\dagger_{xy,\uparrow} d^\dagger_{yz,\downarrow}+2 i d^\dagger_{xy,\downarrow} d^\dagger_{yz,\uparrow}+3 d^\dagger_{yz,\downarrow} d^\dagger_{zx,\downarrow}-3 d^\dagger_{yz,\uparrow} d^\dagger_{zx,\uparrow}\right)},
\end{aligned}
\end{equation}
\begin{equation}
\begin{aligned}
\hat{Y}&=-\frac{1}{4 \sqrt{15}}{i d^\dagger_{z^2,\uparrow} d^\dagger_{x^2-y^2,\uparrow} \left(3 d^\dagger_{yz,\uparrow} d^\dagger_{zx,\downarrow}+3 d^\dagger_{yz,\downarrow} d^\dagger_{zx,\uparrow}-2 d^\dagger_{zx,\downarrow} d^\dagger_{xy,\downarrow}+6 i d^\dagger_{xy,\uparrow} d^\dagger_{yz,\uparrow}-6 d^\dagger_{zx,\uparrow} d^\dagger_{xy,\uparrow}\right)}\\
&+\frac{1}{4 \sqrt{15}}{d^\dagger_{z^2,\downarrow} d^\dagger_{x^2-y^2,\downarrow} \left(-6 d^\dagger_{xy,\downarrow} d^\dagger_{yz,\downarrow}-i \left(3 d^\dagger_{yz,\uparrow} d^\dagger_{zx,\downarrow}+3 d^\dagger_{yz,\downarrow} d^\dagger_{zx,\uparrow}-6 d^\dagger_{zx,\downarrow} d^\dagger_{xy,\downarrow}-2 d^\dagger_{zx,\uparrow} d^\dagger_{xy,\uparrow}\right)\right)}\\
&-\frac{1}{4 \sqrt{15}}{i d^\dagger_{z^2,\uparrow} d^\dagger_{x^2-y^2,\downarrow} \left(-2 d^\dagger_{zx,\uparrow} d^\dagger_{xy,\downarrow}-2 d^\dagger_{zx,\downarrow} d^\dagger_{xy,\uparrow}+3 d^\dagger_{yz,\downarrow} d^\dagger_{zx,\downarrow}+3 d^\dagger_{yz,\uparrow} d^\dagger_{zx,\uparrow}\right)}\\
&-\frac{1}{4 \sqrt{15}}{i d^\dagger_{z^2,\downarrow} d^\dagger_{x^2-y^2,\uparrow} \left(-2 d^\dagger_{zx,\uparrow} d^\dagger_{xy,\downarrow}-2 d^\dagger_{zx,\downarrow} d^\dagger_{xy,\uparrow}+3 d^\dagger_{yz,\downarrow} d^\dagger_{zx,\downarrow}+3 d^\dagger_{yz,\uparrow} d^\dagger_{zx,\uparrow}\right)},
\end{aligned}
\end{equation}
\begin{equation}
\begin{aligned}
\hat{Z}&=\frac{1}{4 \sqrt{15}}{d^\dagger_{z^2,\uparrow} d^\dagger_{x^2-y^2,\uparrow} \left(3 d^\dagger_{zx,\uparrow} d^\dagger_{xy,\downarrow}+3 d^\dagger_{zx,\downarrow} d^\dagger_{xy,\uparrow}-3 i d^\dagger_{xy,\uparrow} d^\dagger_{yz,\downarrow}-3 i d^\dagger_{xy,\downarrow} d^\dagger_{yz,\uparrow}-4 d^\dagger_{yz,\downarrow} d^\dagger_{zx,\downarrow}\right)}\\
&+\frac{1}{4 \sqrt{15}}{d^\dagger_{z^2,\uparrow} d^\dagger_{x^2-y^2,\downarrow} \left(3 d^\dagger_{zx,\downarrow} d^\dagger_{xy,\downarrow}+3 d^\dagger_{zx,\uparrow} d^\dagger_{xy,\uparrow}+3 i d^\dagger_{xy,\downarrow} d^\dagger_{yz,\downarrow}-3 i d^\dagger_{xy,\uparrow} d^\dagger_{yz,\uparrow}-4 d^\dagger_{yz,\uparrow} d^\dagger_{zx,\downarrow}-4 d^\dagger_{yz,\downarrow} d^\dagger_{zx,\uparrow}\right)}\\
&+\frac{1}{4 \sqrt{15}}{d^\dagger_{z^2,\downarrow} d^\dagger_{x^2-y^2,\uparrow} \left(3 d^\dagger_{zx,\downarrow} d^\dagger_{xy,\downarrow}+3 d^\dagger_{zx,\uparrow} d^\dagger_{xy,\uparrow}+3 i d^\dagger_{xy,\downarrow} d^\dagger_{yz,\downarrow}-3 i d^\dagger_{xy,\uparrow} d^\dagger_{yz,\uparrow}-4 d^\dagger_{yz,\uparrow} d^\dagger_{zx,\downarrow}-4 d^\dagger_{yz,\downarrow} d^\dagger_{zx,\uparrow}\right)}\\
&+\frac{1}{4 \sqrt{15}}{d^\dagger_{z^2,\downarrow} d^\dagger_{x^2-y^2,\downarrow} \left(3 d^\dagger_{zx,\uparrow} d^\dagger_{xy,\downarrow}+3 d^\dagger_{zx,\downarrow} d^\dagger_{xy,\uparrow}+3 i d^\dagger_{xy,\uparrow} d^\dagger_{yz,\downarrow}+3 i d^\dagger_{xy,\downarrow} d^\dagger_{yz,\uparrow}-4 d^\dagger_{yz,\uparrow} d^\dagger_{zx,\uparrow}\right)}.
\end{aligned}
\end{equation}
A generic state in the threefold $J=1$ space can be expressed as
\begin{equation}
    \ket{\psi}=b_x\ket{x}+b_y\ket{y}+b_z\ket{z},
\end{equation}
where $\bm{b}=(b_x,b_y,b_z)$ is a complex valued vector that satisfies the normalization $\bm{b}^*\cdot\bm{b}=1$.

\subsection{\texorpdfstring{$d^7=t_{2g}^5e_{g}^2$}{} configuration}
For $d^7$ electron configuration, there are 2 electrons in $e_{g}$ orbitals and 5 electrons in $t_{2g}$ orbitals. Thus, we can denote $d^7$ as $t_{2g}^5e_{g}^2$. There are 3 possibilities to arrange the 5 electrons in $t_{2g}$ orbitals, giving rise to threefold orbital states. In the hole representation, threefold orbital states are given by
\begin{equation}
\begin{split}
& \ket{d}=D^\dagger|0\rangle=d^\dagger_{3z^2-r^2}d^\dagger_{x^2-y^2}d_{xy}^\dagger\ket{0}, \\
& \ket{e}=E^\dagger|0\rangle=d^\dagger_{3z^2-r^2}d^\dagger_{x^2-y^2}d_{yz}^\dagger\ket{0}, 
\\
& \ket{f}=F^\dagger|0\rangle=d^\dagger_{3z^2-r^2}d^\dagger_{x^2-y^2}d_{zx}^\dagger \ket{0}.
\end{split}
\end{equation}
The projection of the physical orbital angular momentum $\bm{L}_{\text{physical}}$ on the threefold orbital states gives an effective angular momentum $\bm{L}\cong-\bm{L}_{\text{physical}}$ with the quantum number $L=1$. The eigenstates of $L_z$ are
\begin{equation}\label{eq:d7_L=1}
\begin{split}
& \ket{L_z=+1}=\frac{1}{\sqrt{2}}(\ket{e}+i\ket{f}), \\
& \ket{L_z=0}=\ket{d}, \\
& \ket{L_z=-1}=\frac{1}{\sqrt{2}}(\ket{e}-i\ket{f}).
\end{split}
\end{equation}
The first Hund's rule tells us the $d^7$ configuration has a total spin $\bm{S}$ with the spin quantum number $S=3/2$. The eigenstates of $S_z$ are given by
\begin{equation}\label{eq:d7_S=3/2}
\begin{split}
& \ket{S_z=3/2}=\ket{\uparrow\uparrow\uparrow}, \\
& \ket{S_z=1/2}=\frac{1}{\sqrt{3}}\left(\ket{\uparrow\uparrow\downarrow}+\ket{\uparrow\uparrow\downarrow}+\ket{\downarrow\uparrow\uparrow}\right), \\
& \ket{S_z=-1/2}=\frac{1}{\sqrt{3}}\left(\ket{\downarrow\downarrow\uparrow}+\ket{\downarrow\uparrow\downarrow}+\ket{\uparrow\downarrow\downarrow}\right), \\
& \ket{S_z=-3/2}=\ket{\downarrow\downarrow\downarrow}.
\end{split}
\end{equation}
After the SOC, the total angular momentum $\bm{J}$ of the ground states has quantum number $J=1/2$. These states are given by
\begin{equation}\label{eq:d7_J=1/2}
\begin{split}
& \ket{J_z=1/2}=\phi_{1/2}^\dagger\ket{0}=\sqrt{\frac{1}{6}}\ket{L_z=1,S_z=-1/2}-\sqrt{\frac{1}{3}}\ket{L_z=0,S_z=1/2}+\sqrt{\frac{1}{2}}\ket{L_z=-1,S_z=3/2}, \\
& \ket{J_z=-1/2}=\phi_{-1/2}^\dagger\ket{0}=\sqrt{\frac{1}{2}}\ket{L_z=1,S_z=-3/2}-\sqrt{\frac{1}{3}}\ket{L_z=0,S_z=-1/2}+\sqrt{\frac{1}{6}}\ket{L_z=-1,S_z=1/2}.
\end{split}
\end{equation}
Using \cref{eq:d7_L=1,eq:d7_S=3/2}, we can write \cref{eq:d7_J=1/2} in terms of hole operators
\begin{equation}
\begin{aligned}
    \phi_{1/2}^\dagger=&-\frac{1}{3} d^\dagger_{z^2,\uparrow} d^\dagger_{x^2-y^2,\downarrow} d^\dagger_{xy,\downarrow}-\frac{1}{3} d^\dagger_{z^2,\downarrow} d^\dagger_{x^2-y^2,\uparrow} d^\dagger_{xy,\downarrow}-\frac{1}{3} d^\dagger_{z^2,\downarrow} d^\dagger_{x^2-y^2,\downarrow} d^\dagger_{xy,\uparrow}+\frac{1}{6} d^\dagger_{z^2,\uparrow} d^\dagger_{x^2-y^2,\uparrow} d^\dagger_{yz,\downarrow}\\
    &+\frac{1}{6} d^\dagger_{z^2,\uparrow} d^\dagger_{x^2-y^2,\downarrow} d^\dagger_{yz,\uparrow}+\frac{1}{6} d^\dagger_{z^2,\downarrow} d^\dagger_{x^2-y^2,\uparrow} d^\dagger_{yz,\uparrow}+\frac{1}{6} i d^\dagger_{z^2,\uparrow} d^\dagger_{x^2-y^2,\uparrow} d^\dagger_{zx,\downarrow}+\frac{1}{6} i d^\dagger_{z^2,\uparrow} d^\dagger_{x^2-y^2,\downarrow} d^\dagger_{zx,\uparrow}\\
    &+\frac{1}{6} i d^\dagger_{z^2,\downarrow} d^\dagger_{x^2-y^2,\uparrow} d^\dagger_{zx,\uparrow}+\frac{1}{2} d^\dagger_{z^2,\downarrow} d^\dagger_{x^2-y^2,\downarrow} d^\dagger_{yz,\downarrow}-\frac{1}{2} i d^\dagger_{z^2,\downarrow} d^\dagger_{x^2-y^2,\downarrow} d^\dagger_{zx,\downarrow},
\end{aligned}    
\end{equation}
\begin{equation}
    \begin{aligned}
        \phi_{-1/2}^\dagger=&-\frac{1}{3} d^\dagger_{z^2,\uparrow} d^\dagger_{x^2-y^2,\downarrow} d^\dagger_{xy,\downarrow}-\frac{1}{3} d^\dagger_{z^2,\downarrow} d^\dagger_{x^2-y^2,\uparrow} d^\dagger_{xy,\downarrow}-\frac{1}{3} d^\dagger_{z^2,\downarrow} d^\dagger_{x^2-y^2,\downarrow} d^\dagger_{xy,\uparrow}+\frac{1}{6} d^\dagger_{z^2,\uparrow} d^\dagger_{x^2-y^2,\uparrow} d^\dagger_{yz,\downarrow}\\
        &+\frac{1}{6} d^\dagger_{z^2,\uparrow} d^\dagger_{x^2-y^2,\downarrow} d^\dagger_{yz,\uparrow}+\frac{1}{6} d^\dagger_{z^2,\downarrow} d^\dagger_{x^2-y^2,\uparrow} d^\dagger_{yz,\uparrow}+\frac{1}{6} i d^\dagger_{z^2,\uparrow} d^\dagger_{x^2-y^2,\uparrow} d^\dagger_{zx,\downarrow}+\frac{1}{6} i d^\dagger_{z^2,\uparrow} d^\dagger_{x^2-y^2,\downarrow} d^\dagger_{zx,\uparrow}\\
        &+\frac{1}{6} i d^\dagger_{z^2,\downarrow} d^\dagger_{x^2-y^2,\uparrow} d^\dagger_{zx,\uparrow}+\frac{1}{2} d^\dagger_{z^2,\downarrow} d^\dagger_{x^2-y^2,\downarrow} d^\dagger_{yz,\downarrow}-\frac{1}{2} i d^\dagger_{z^2,\downarrow} d^\dagger_{x^2-y^2,\downarrow} d^\dagger_{zx,\downarrow}.
    \end{aligned}
\end{equation}

\subsection{\texorpdfstring{$d^7=t_{2g}^4e_{g}^3$}{} configuration}
After hopping, we can have $d^7$ electron configuration as intermediate states, with 3 electrons in $e_{g}$ orbitals and 4 electrons in $t_{2g}$ orbitals. 
Thus, we can denote $d^7$ as $t_{2g}^4e_{g}^3$.
There are 3 possibilities to arrange the 4 electrons in $t_{2g}$ orbitals, giving rise to threefold orbital states. 
And 2 possibilities for 3 electrons in $e_g$ orbitals.
In the hole representation, threefold orbital states are also given by
\begin{equation}
\begin{split}
& \ket{g}=G^\dagger|0\rangle=
d^\dagger_{e_g}d^\dagger_{zx}d_{xy}^\dagger\ket{0}, \\
& \ket{h}=H^\dagger|0\rangle=
d^\dagger_{e_g}d^\dagger_{xy}d_{yz}^\dagger\ket{0}, 
\\
& \ket{j}=J^\dagger|0\rangle=
d^\dagger_{e_g}d^\dagger_{yz}d_{zx}^\dagger \ket{0},
\end{split}
\end{equation}
where $e_g$ represents one of the $e_g$ orbitals, either $3z^2-r^2$ or $x^2-y^2$.
The projection of the physical orbital angular momentum $\bm{L}_{\text{physical}}$ on the threefold orbital states gives an effective angular momentum $\bm{L}\cong-\bm{L}_{\text{physical}}$ with the quantum number $L=1$. The eigenstates of $L_z$ are similar to the Eqs.~\eqref{eq:d7_L=1}.
The first Hund's rule tells us the $d^7$ configuration has a total spin $\bm{S}$ with the spin quantum number $S=3/2$. The eigenstates of $S_z$ are similar to Eqs.~\eqref{eq:d7_S=3/2}.
After the SOC, the total angular momentum $\bm{J}$ of the ground states has quantum number $J=1/2$.
These states are similar to the Eqs.~\eqref{eq:d7_J=1/2}.
Therefore, we could obtain the ground state in terms of hole operators,

\begin{equation}
\begin{aligned}
\phi_{1/2}^\dagger=&
-\frac{1}{3} d^\dagger_{e_g,\uparrow} d^\dagger_{zx,\downarrow} d^\dagger_{xy,\downarrow}
-\frac{1}{3} d^\dagger_{e_g,\downarrow} d^\dagger_{zx,\uparrow} d^\dagger_{xy,\downarrow}
-\frac{1}{3} d^\dagger_{e_g,\downarrow} d^\dagger_{zx,\downarrow} d^\dagger_{xy,\uparrow}
+\frac{1}{6} d^\dagger_{e_g,\uparrow} d^\dagger_{xy,\uparrow} d^\dagger_{yz,\downarrow}\\
&+\frac{1}{6} d^\dagger_{e_g,\uparrow} d^\dagger_{xy,\downarrow} d^\dagger_{yz,\uparrow}
+\frac{1}{6} d^\dagger_{e_g,\downarrow} d^\dagger_{xy,\uparrow} d^\dagger_{yz,\uparrow}
+\frac{1}{6} i d^\dagger_{e_g,\uparrow} d^\dagger_{yz,\uparrow} d^\dagger_{zx,\downarrow}
+\frac{1}{6} i d^\dagger_{e_g,\uparrow} d^\dagger_{yz,\downarrow} d^\dagger_{zx,\uparrow}\\
&+\frac{1}{6} i d^\dagger_{e_g,\downarrow} d^\dagger_{yz,\uparrow} d^\dagger_{zx,\uparrow}
+\frac{1}{2} d^\dagger_{e_g,\downarrow} d^\dagger_{xy,\downarrow} d^\dagger_{yz,\downarrow}
-\frac{1}{2} i d^\dagger_{e_g,\downarrow} d^\dagger_{yz,\downarrow} d^\dagger_{zx,\downarrow},
\end{aligned}    
\end{equation}
\begin{equation}
\begin{aligned}
\phi_{-1/2}^\dagger=&
-\frac{1}{3} d^\dagger_{e_g,\uparrow} d^\dagger_{zx,\downarrow} d^\dagger_{xy,\downarrow}
-\frac{1}{3} d^\dagger_{e_g,\downarrow} d^\dagger_{zx,\uparrow} d^\dagger_{xy,\downarrow}
-\frac{1}{3} d^\dagger_{e_g,\downarrow} d^\dagger_{zx,\downarrow} d^\dagger_{xy,\uparrow}
+\frac{1}{6} d^\dagger_{e_g,\uparrow} d^\dagger_{xy,\uparrow} d^\dagger_{yz,\downarrow}\\
&
+\frac{1}{6} d^\dagger_{e_g,\uparrow} d^\dagger_{xy,\downarrow} d^\dagger_{yz,\uparrow}
+\frac{1}{6} d^\dagger_{e_g,\downarrow} d^\dagger_{xy,\uparrow} d^\dagger_{yz,\uparrow}
+\frac{1}{6} i d^\dagger_{e_g,\uparrow} d^\dagger_{yz,\uparrow} d^\dagger_{zx,\downarrow}
+\frac{1}{6} i d^\dagger_{e_g,\uparrow} d^\dagger_{yz,\downarrow} d^\dagger_{zx,\uparrow}\\
&
+\frac{1}{6} i d^\dagger_{e_g,\downarrow} d^\dagger_{yz,\uparrow} d^\dagger_{zx,\uparrow}
+\frac{1}{2} d^\dagger_{e_g,\downarrow} d^\dagger_{xy,\downarrow} d^\dagger_{yz,\downarrow}-\frac{1}{2} i d^\dagger_{e_g,\downarrow} d^\dagger_{yz,\downarrow} d^\dagger_{zx,\downarrow}.
\end{aligned}
\end{equation}

\subsection{\texorpdfstring{$d^8=t_{2g}^6e_{g}^2$}{} configuration}
After hopping, we can have $d^8$ electron configuration as the intermediate state, with 2 electrons in $e_{g}$ orbitals and 6 electrons in $t_{2g}$ orbitals. Thus, we can denote $d^8$ as $t_{2g}^6e_{g}^2$. There is only one possibility to arrange these electrons, giving rise to an orbital state
\begin{equation}
    \ket{L_z=0}=d_{3z^2-r^2}^\dagger d_{x^2-y^2}^\dagger\ket{0}.
\end{equation}
According to Hund's rule, the total spin momentum $\bm{S}$ has quantum number $S=1$. The eigenstates of $S_z$ are
\begin{equation}
\begin{split}
& \ket{S_z=1}=\ket{\uparrow\uparrow}, \\
& \ket{S_z=0}=\frac{1}{\sqrt{2}}\left(\ket{\uparrow\downarrow}+\ket{\downarrow\uparrow}\right), \\
& \ket{S_z=-1}=\ket{\downarrow\downarrow}.
\end{split}
\label{eq:S=1}
\end{equation}
After the SOC, the total angular momentum $\bm{J}$ as quantum number $J=1$ and the eigenstates are $\ket{J_z=j_z}=\varphi_{j_z}^\dagger\ket{0}$ ($j_z=1,0,-1$) with
\begin{equation}
    \varphi_{1}^\dagger=d_{3z^2-r^2,\uparrow}^\dagger d_{x^2-y^2,\uparrow}^\dagger,
\end{equation}
\begin{equation}
    \varphi_0^\dagger=\frac{1}{\sqrt{2}}\left(d_{3z^2-r^2,\uparrow}^\dagger d_{x^2-y^2,\downarrow}^\dagger+d_{3z^2-r^2,\downarrow}^\dagger d_{x^2-y^2,\uparrow}^\dagger\right),
\end{equation}
\begin{equation}
    \varphi_{-1}^\dagger=d_{3z^2-r^2,\downarrow}^\dagger d_{x^2-y^2,\downarrow}^\dagger.
\end{equation}

\subsection{\texorpdfstring{$d^8=t_{2g}^5e_{g}^3$}{} configuration}
After hopping, we can also have $d^8$ electron configuration as the intermediate state, with 3 electrons in $e_{g}$ orbitals and 5 electrons in $t_{2g}$ orbitals. Thus, we can denote $d^8$ as $t_{2g}^5e_{g}^3$. 
There are three possible states to arrange 5 electrons in $t_{2g}$ orbitals and two possible states for 3 electrons in $e_{g}$ orbitals. 
In the hole representation, we have,
\begin{equation}
\begin{split}
& \ket{k}=K^\dagger|0\rangle=
d^\dagger_{e_g} d_{xy}^\dagger\ket{0}, 
\\
& \ket{p}=P^\dagger|0\rangle=
d^\dagger_{e_g} d_{yz}^\dagger\ket{0}, 
\\
& \ket{r}=R^\dagger|0\rangle=
d^\dagger_{e_g} d_{zx}^\dagger\ket{0},
\end{split}
\end{equation}
where $e_g$ represents one of the $e_g$ orbitals, either $3z^2-r^2$ or $x^2-y^2$.
The projection of the physical orbital angular momentum $\bm{L}_{\text{physical}}$ on the threefold orbital states gives an effective angular momentum $\bm{L}\cong-\bm{L}_{\text{physical}}$ with the quantum number $L=1$. The eigenstates of $L_z$ are similar to the Eqs.~\eqref{eq:d7_L=1}.
According to Hund's rule, the total spin angular momentum $\bm{S}$ has quantum number $S=1$ with eigen states Eq.~\eqref{eq:S=1}. 
After the SOC, the total angular momentum $\bm{J}$ has quantum number $J=0$ and the only eigenstate is 
\begin{equation}
\begin{aligned}
|J=0\rangle
=\varphi^\dagger_0\ket{0}
&=
\frac{1}{3}\bigg( 
\ket{L_z=1,wS_z=-1} - \ket{L_z=0,S_z=0} +\ket{L_z=-1,S_z=1}
\bigg)
\\&=
\frac{1}{3\sqrt{2}}
\bigg(
d_{e_g\downarrow}^\dagger d_{yz\downarrow}^\dagger
+ i d_{e_g\downarrow}^\dagger d_{zx\downarrow}^\dagger
- d_{e_g\downarrow}^\dagger d_{xy\uparrow}^\dagger
- d_{e_g\uparrow}^\dagger d_{xy\downarrow}^\dagger
+ d_{e_g\uparrow}^\dagger d_{yz\uparrow}^\dagger
-i d_{e_g\uparrow}^\dagger d_{zx\downarrow}^\dagger
\bigg)
\ket{0}
\end{aligned}
\end{equation}

\section{Single site Hamiltonian and magnetic moments}
\label{app:single site Ham}
Given a system with a total angular momentum $\bm{J}$, the generic Hamiltonian should be a function of $\bm{J}$. Expanding the Hamiltonian in terms of polynomials of $\bm{J}$, we obtain
\begin{equation}
    H_0=\sum_\mu a_\mu J_\mu + \sum_{\mu\nu} b_{\mu\nu} J_{\mu} J_\nu + \sum_{\mu\nu\sigma} c_{\mu\nu\sigma} J_\mu J_\nu J_\sigma + \cdots.
\end{equation}
It is clear that $H_0$ is diagonal for different angular momentum quantum numbers $J$. For the cases considered in this paper, the largest $J$ is $1$. Thus, polynomials with more than 2 $J_\mu$ either are constant or can be absorbed into polynomials with lower order. We can truncate $H_0$ at the 2nd order and obtain
\begin{equation}
    H_0=\sum_\mu a_\mu J_\mu + \sum_{\mu\nu} b_{\mu\nu} J_{\mu} J_\nu.
\end{equation}
For generic $a_\mu$ and $b_{\mu\nu}$, the eigenstates of $H_0$ are hard to obtain. Therefore, we focus on a simpler case, in which
\begin{equation}
    H_0=J_1\hat{\bm{e}}\cdot\bm{J}+J_2\left(\hat{\bm{e}}\cdot\bm{J}\right)^2,
\end{equation}
where $\hat{\bm{e}}=(\sin\theta\cos\phi,\sin\theta\sin\phi,\cos\theta)$ is a unit vector. The eigenstates of $H_0$ are these of $\hat{\bm{e}}\cdot\bm{J}$. In the $J=1/2$ subspace, eigenstates are
\begin{equation}
\begin{split}
& \ket{\hat{\bm{e}}\cdot\bm{J}=1/2}=\cos\frac{\theta}{2}\ket{\Uparrow}+e^{i\phi}\sin\frac{\theta}{2}\ket{\Downarrow},\quad \text{with energy $\frac{J_2+2J_1}{4}$},\\
& \ket{\hat{\bm{e}}\cdot\bm{J}=-1/2}=\sin\frac{\theta}{2}\ket{\Uparrow}-e^{i\phi}\cos\frac{\theta}{2}\ket{\Downarrow},\quad \text{with energy $\frac{J_2-2J_1}{4}$},
\end{split}
\end{equation}
where we have defined $\ket{\Uparrow}=\ket{J_z=1/2}$ and $\ket{\Downarrow}=\ket{J_z=-1/2}$. In the $J=1$ subspace, eigenstates are
\begin{equation}
\begin{split}
& \ket{\hat{\bm{e}}\cdot\bm{J}=1}=\frac{1}{\sqrt{2}}\left[(\cos\theta\cos\phi-i\sin\phi)\ket{x}+(\cos\theta\sin\phi+i\cos\phi)\ket{y}-\sin\theta\ket{z}\right],\quad \text{with energy $J_2+J_1$},\\
& \ket{\hat{\bm{e}}\cdot\bm{J}=0}=\sin\theta\cos\phi\ket{x}+\sin\theta\sin\phi\ket{y}+\cos\theta\ket{z},\quad \text{with energy $0$},\\
& \ket{\hat{\bm{e}}\cdot\bm{J}=-1}=\frac{1}{\sqrt{2}}\left[(\cos\theta\cos\phi+i\sin\phi)\ket{x}+(\cos\theta\sin\phi-i\cos\phi)\ket{y}-\sin\theta\ket{z}\right],\quad \text{with energy $J_2-J_1$}.\\
\end{split}
\end{equation}
Without loss of generality, we assume $J_1> 0$ and $J_2> 0$. In this case the ground state of $H_0$ in the $J=1/2$ subspace is always $\ket{\hat{\bm{e}}\cdot\bm{J}=-1/2}$. In the $J=1$ subspace, the ground state is $\ket{\hat{\bm{e}}\cdot\bm{J}=0}$ if $J_2>J_1$, and is $\ket{\hat{\bm{e}}\cdot\bm{J}=-1}$ if $J_2<J_1$.

We are interested in the magnetic moments carried by these states. For a generic state in the $J=1/2$ space,
\begin{equation}
    \ket{\phi}=a_{\Uparrow}\ket{\Uparrow}+a_{\Downarrow}\ket{\Downarrow},
\end{equation}
one can calculate its magnetic dipole
\begin{equation}
\begin{split}
& \braket{\phi|J_x|\phi}=a_\Uparrow^*a_\Downarrow+a_\Downarrow^*a_\Uparrow, \\
& \braket{\phi|J_y|\phi}=-i\left(a_\Uparrow^*a_\Downarrow-a_\Downarrow^*a_\Uparrow\right), \\
& \braket{\phi|J_z|\phi}=a_\Uparrow^*a_\Uparrow-a_\Downarrow^*a_\Downarrow. \\
\end{split}
\end{equation}
If we choose $\ket{\phi}$ to be the ground state $\ket{\hat{\bm{e}}\cdot\bm{J}=-1/2}$, then $\braket{\bm{J}}=-\hat{\bm{e}}$. Namely, the magnetic dipole is polarized along $-\hat{\bm{e}}$. For a generic state in the $J=1$ space,
\begin{equation}
    \ket{\psi}=b_x\ket{x}+b_y\ket{y}+b_z\ket{z},
\end{equation}
both magnetic dipole and quadrupole appear. The magnetic dipole is
\begin{equation}
    \braket{\psi|\bm{J}|\psi}=-i\bm{b}^*\times\bm{b}.
\end{equation}
One can find that the magnetic dipole for the unpolarized state $\ket{\hat{\bm{e}}\cdot\bm{J}=0}$ vanished, while equals $-\hat{\bm{e}}$ for the polarized state $\ket{\hat{\bm{e}}\cdot\bm{J}=-1}$. The magnetic quadrupole is defined as
\begin{equation}
    Q_{\mu\nu}=\frac{1}{2}\left(J_\mu J_\nu + J_\nu J_\mu\right)-\frac{\bm{J}^2}{3}\delta_{\mu\nu}.
\end{equation}
Its expectation value on $\ket{\psi}$ can be obtained
\begin{equation}
    \braket{\psi|Q_{\mu\nu}|\psi}=\frac{1}{3}\delta_{\mu\nu}-\frac{1}{2}\left(b_\mu^* b_\nu + b_\nu^* b_\mu\right).
\end{equation}
Although the tensor $Q_{\mu\nu}$ has 9 components, only 5 components are independent. They are,
\begin{equation}
\begin{split}
& \braket{\psi|Q_{3z^3-r^2}|\psi}=-\sqrt{3}\left(\braket{\psi|Q_{xx}|\psi}+\braket{\psi|Q_{yy}|\psi}\right)=-\frac{b_x^*b_x+b_y^*b_y}{\sqrt{3}}, \\
& \braket{\psi|Q_{x^2-y^2}|\psi}=\braket{\psi|Q_{xx}|\psi}-\braket{\psi|Q_{yy}|\psi}=-(b_x^*b_x-b_y^*b_y), \\
& \braket{\psi|Q_{xy}|\psi}=-\frac{1}{2}(b_x^*b_y+b_y^*b_x), \\
& \braket{\psi|Q_{yz}|\psi}=-\frac{1}{2}(b_y^*b_z+b_z^*b_y), \\
& \braket{\psi|Q_{zx}|\psi}=-\frac{1}{2}(b_z^*b_x+b_x^*b_z).
\end{split}
\end{equation}

\section{Full equations of electric polarization }
\label{app:Full P}
Here we consider a generic $J_1$ and $J_2$ for the single site Hamiltonian $H_0$, leading to a general expression for $\bm b$ with complex components.
For $d^7$ electron state, $\bm a=(\sin \theta/2, -e^{i\phi}\cos\theta/2)$ is considered to be a complex vector under the effective ordering field $\hat{\bm{e}}$.
Then under this condition, the ground state has both dipole and quadrupole order in $d^6$ electron state, and dipole-only order $d^7$ electron state.
The electric polarization $\bm{P}$ contains onsite contributions and mixed-site contributions. Due to a model-dependent anisotropic factor $s_\mu$, we may define a scaled polarization $\tilde{P}_\mu=P_\mu/s_\mu$. The onsite contributions are given by
\begin{equation}
\begin{split}
\tilde{P}^x_{\text{on}}&=\left(24|b_{1x}|^2+25|b_{1y}|^2\right)-\left(24|b_{2x}|^2+25|b_{2y}|^2\right)
\\&\quad+
3(\sqrt{3}-1)r_t\bigg(b_{1y} b_{1x}^*+b_{1x} b_{1y}^*-i b_{1z} b_{1x}^*+i b_{1x} b_{1z}^*+i b_{1z} b_{1y}^*-i b_{1y} b_{1z}^*
\\&\quad
-b_{2y} b_{2x}^*-b_{2x} b_{2y}^*+i b_{2z} b_{2x}^*-i b_{2x} b_{2z}^*-i b_{2z} b_{2y}^*+i b_{2y} b_{2z}^*\bigg),
\end{split}
\end{equation}
\begin{equation}
    \tilde{P}^y_{\text{on}}= \left(b_{1x}^*b_{1y}+b_{1y}^*b_{1x}\right)-\left(b_{2x}^*b_{2y}+b_{2y}^*b_{2x}\right),
\end{equation}
\begin{equation}
\tilde{P}^z_{\text{on}}=
\left[-i\left(b_{1y}^*b_{1z}-b_{1y}b_{1z}^*\right)+i\left(b_{2y}^*b_{2z}-b_{2y}b_{2z}^*\right)\right]-8\left[\left(a_{1\Uparrow}^*a_{1\Downarrow}+a_{1\Downarrow}^*a_{1\Uparrow}\right)-\left(a_{2\Uparrow}^*a_{2\Downarrow}+a_{2\Downarrow}^*a_{2\Uparrow}\right)\right],
\end{equation}
where $r_t=\frac{t_0}{t_1}$, $s_x=4C$, $s_y=2C$, and $s_z=10C$. We have defined a constant $C=e t_1 I/(480 \Delta)$, and $I$ is the following integral
\begin{equation}
    I=\int\diff^3r\,d_{xy}(\bm{r})xp_y(\bm{r})
\end{equation}
and its cyclic permutation in $x,y,z$. Onsite contributions can be expressed in terms of magnetic dipole and quadrupole,
\begin{equation}
\begin{split}
\tilde{P}_{\text{on}}^x&=
\bigg[\frac{49}{\sqrt{3}}\braket{Q_{3z^2-r^2}}_1+
\braket{Q_{x^2-y^2}}_1
-3(\sqrt{3}-1)(2\braket{Q_{xy}}_1
+\braket{J_x}_1^{(d^6)}+\braket{J_y}_1^{(d^6)})\bigg]
\\&-
\bigg[\frac{49}{\sqrt{3}}\braket{Q_{3z^2-r^2}}_2+
\braket{Q_{x^2-y^2}}_2
-3(\sqrt{3}-1)(2\braket{Q_{xy}}_2
+\braket{J_x}_2^{(d^6)}+\braket{J_y}_2^{(d^6)})\bigg],
\\
\tilde{P}_{\text{on}}^y&=-\bigg(\braket{Q_{xy}}_1-\braket{Q_{xy}}_2\bigg), 
\\ \tilde{P}_{\text{on}}^z&=
\bigg[
\braket{J_x}_1^{(d^6)}-8\braket{J_x}_1^{(d^7)}
\bigg]
-
\bigg[
\braket{J_x}_2^{(d^6)}-8\braket{J_x}_2^{(d^7)}
\bigg],
\end{split}
\end{equation}

The mixed-site  contributions are written as
\begin{equation}
    \tilde{P}^x_{\text{mix}}=0,
\end{equation}
\begin{equation}
\begin{aligned}
    \tilde{P}^y_{\text{mix}}&=a_{1\Downarrow}  a_{2\Downarrow}^* (b_{2y}   b_{1x}^*-b_{2x}   b_{1y}^*+2 i b_{2z}   b_{1z}^*)+ a_{2\Downarrow}  a_{1\Downarrow}^* (b_{1x}   b_{2y}^*-b_{1y}   b_{2x}^*-2 i b_{1z}   b_{2z}^*)\\
    &+ a_{1\Downarrow}  a_{2\Uparrow}^* (i b_{2z}   b_{1x}^*+b_{2z}   b_{1y}^*+i b_{2x}   b_{1z}^*+b_{2y}   b_{1z}^*)+ a_{2\Uparrow}  a_{1\Downarrow}^* (-i b_{1x}   b_{2z}^*+b_{1y}   b_{2z}^*-i b_{1z}   b_{2x}^*+b_{1z}   b_{2y}^*)\\
    &+ a_{1\Uparrow}  a_{2\Downarrow}^* (i b_{2z}   b_{1x}^*-b_{2z}   b_{1y}^*+i b_{2x}   b_{1z}^*-b_{2y}   b_{1z}^*)+ a_{2\Downarrow}  a_{1\Uparrow}^* (-i b_{1x}   b_{2z}^*-b_{1y}   b_{2z}^*-i b_{1z}   b_{2x}^*-b_{1z}   b_{2y}^*)\\
    &+ a_{1\Uparrow}  a_{2\Uparrow}^* (b_{2y}   b_{1x}^*-b_{2x}   b_{1y}^*-2 i b_{2z}   b_{1z}^*)+ a_{2\Uparrow}  a_{1\Uparrow}^* (b_{1x}   b_{2y}^*-b_{1y}   b_{2x}^*+2 i b_{1z}   b_{2z}^*),
\end{aligned}
\end{equation}
\begin{equation}
\begin{aligned}
    \tilde{P}^z_{\text{mix}} &= a_{1\Downarrow} a_{2\Downarrow}^* (-b_{2z}   b_{1x}^*+i b_{2z}   b_{1y}^*+b_{2x}   b_{1z}^*+i b_{2y}   b_{1z}^*)+a_{2\Downarrow} a_{1\Downarrow}^* (-b_{1x}   b_{2z}^*-i b_{1y}   b_{2z}^*+b_{1z}   b_{2x}^*-i b_{1z}   b_{2y}^*)\\
    &+a_{1\Downarrow} a_{2\Uparrow}^* (i b_{2x}   b_{1y}^*-i b_{2y}   b_{1x}^*)+a_{2\Uparrow} a_{1\Downarrow}^* (i b_{1x}   b_{2y}^*-i b_{1y}   b_{2x}^*)\\
    &+a_{1\Uparrow} a_{2\Downarrow}^* (i b_{2y}   b_{1x}^*-i b_{2x}   b_{1y}^*)+a_{2\Downarrow} a_{1\Uparrow}^* (i b_{1y}   b_{2x}^*-i b_{1x}   b_{2y}^*)\\
    &+a_{1\Uparrow} a_{2\Uparrow}^* (b_{2z}   b_{1x}^*+i b_{2z}   b_{1y}^*-b_{2x}   b_{1z}^*+i b_{2y}   b_{1z}^*)+a_{2\Uparrow} a_{1\Uparrow}^* (b_{1x}   b_{2z}^*-i b_{1y}   b_{2z}^*-b_{1z}   b_{2x}^*-i b_{1z}   b_{2y}^*).
\end{aligned}
\end{equation}

\clearpage
\twocolumngrid


\bibstyle{apsrev-nourl}
\bibliography{ref}

\end{document}